*To be submitted to Ferroelectrics*

# Polarization Screening Effect on Local Polarization Switching Mechanism and Hysteresis Loop Measurements in Piezoresponse Force Microscopy

Anna N. Morozovska,[1,*] Sergei V. Kalinin,[†,‡], Eugene A. Eliseev[3] and Sergei V. Svechnikov[1]

[1]V. Lashkaryov Institute of Semiconductor Physics, National Academy of Science of Ukraine, 41, pr. Nauki, 03028 Kiev, Ukraine

[2]Materials Sciences and Technology Division and the Center for Nanophase Materials Science, Oak Ridge National Laboratory, Oak Ridge, TN 37831

[3]Institute for Problems of Materials Science, National Academy of Science of Ukraine, 3, Krjijanovskogo, 03142 Kiev, Ukraine

Piezoresponse Force Spectroscopy (PFS) has emerged as a powerful tool for probing polarization dynamics on the nanoscale. Application of a dc bias to a nanoscale probe in contact with a ferroelectric surface results in the nucleation and growth of a ferroelectric domain below the probe apex. The latter affects local electromechanical response detected by the probe. Resulting hysteresis loop contains information on local ferroelectric switching. The self-consistent analysis of the PFS data requires (a) deriving the thermodynamic parameters of domain nucleation and (b) establishing the relationships between domain parameters and PFM signal. Here, we analyze the effect of screening at the domain wall on local polarization reversal mechanism. It is shown that the screening control both the domain nucleation activation energy and hysteresis loop saturation rate.

[*] Corresponding author, morozo@i.com.ua
[†] Corresponding author, sergei2@ornl.gov
[‡] Also affiliated with the Department of Materials Science and Engineering, North Carolina State University, Raleigh, NC, and Department of Materials Sciences and Engineering, University of Tennessee, Knoxville.

## 1. Overview

The presence of two or more stable polarization states has propelled ferroelectric materials into a spotlight as promising materials for non-volatile random access memories,[1,2] high-density data storage, and functional oxide heterostructures.[3] Nanoscale ferroelectric domain patterning was proposed as basis for ferroelectric data storage devices with recently demonstrated minimal bit size of ~8 nm, corresponding to storage density of ~10 Tbit/inch$^2$.[4] Polarization dependence of chemical reactivity in acid dissolution[5] or metal photodeposition[6] processes allows domain-based nanofabrication. These new applications, as well as the need for understanding individual domain dynamics in ferroelectric devices, have necessitated local studies of domain growth processes. A number of experimental studies of domain growth kinetics by Piezoresponse Force Microscopy (PFM) have been reported.[7,8,9]. It was shown that domain growth follows an approximately logarithmic dependence on the pulse length and a linear dependence in magnitude.[7] However, the measurements based on domain switching with subsequent imaging of the size of formed domain are extremely time consuming and do not allow studying spatial variability of switching behavior.

Piezoresponse Force Spectroscopy (PFS) has emerged as a powerful technique to probe bias-controlled local polarization dynamics in ferroelectric materials. The probe concentrates an electric field to a nanoscale volume of material (~10-30 nm), and induces local domain formation. Simultaneously, the probe detects the onset of nucleation and the size of a forming domain via detection of the electromechanical response of the material to a small AC bias after the application of controlled DC bias pulses.[10] From these hysteresis loops, phenomenological characteristics such as local imprint, coercive bias, and work of switching can be obtained. This information is important for any technological application based on ferroelectric switching (e.g. FeRAM) and is required for understanding of the fundamental mechanisms of polarization reversal including the role of defects, interfaces, and topography on polarization switching.

PFM applications for domain patterning and spectroscopic studies of polarization dynamics necessitate quantitative analysis of tip-induced polarization switching process, including the size of the nucleating domain and corresponding activation energy. Thermodynamics of domain switching in the Landauer approximation[11] for domain shape and point charge approximation for the tip was given by Abplanalp[12] and independently by Molotskii et al.[13, 14, 15] and Shvebelman.[16] Using the Landauer model[11] and a point-charge approximation for the electric field of AFM tip, they obtained elegant closed-form analytical expressions for the domain size dependence on the applied voltage in the case when the surface charges were completely compensated by the external screening charges. The interaction with the charged AFM probe was calculated as if these screening charges were absent. It was shown by Kalinin et al.[17] that capacitance approximation for the AFM tip electric field is applicable only for large domain sizes, while the description of switching on the length scales comparable to the tip radius of curvature and higher-order switching phenomena, requires exact electroelastic field structure to be taken into account. For realistic tip geometries, domain nucleation requires a certain threshold bias on the order of 0.1-10 V [18], sufficient to nucleate a domain in the finite electric field of the tip. Detailed analysis of polarization switching including nucleation and domain growth in ferroelectrics depending on material parameters such as Debye length, surface screening, and tip geometry using simplified Pade approximations for the free energy has been reported by Morozovska and Eliseev in a series of recent papers.[19, 20]

The 1D model for piezoelectric hysteresis loop formation was developed by Kalinin et al.[21] and later independently by Kholkin.[22,23] Recently proposed self-consistent 3D model [24, 25], [26] analyzes the signal formation mechanism in PFM by deriving the main parameters of domain nucleation in semi-infinite material and establishing the relationships between domain parameters and the PFM signal using linear Green's function theory. Recently, Gerra et al. [27] and Huber [28] extended Landauer model to consider the effects of surface-stimulated nucleation and mechanical constrain respectively on the polarization switching in the homogeneous external field.

Despite this progress, analytical closed form solutions are now available only for the late stages of domain growth process. In this manuscript, we present the analytical description of the early stages of polarization switching, in particular domain nucleation, and discuss the role of internal screening on domain nucleation and its effective piezoelectric response.

## 2. Thermodynamics of domain formation

The thermodynamics of the switching process can be analyzed from the bias dependence of the free energy of the nascent domain:

$$\Phi(r,l,U) = U\Phi_U(r,l) + \Phi_S(r,l) + \Phi_D(r,l), \tag{1}$$

where $\Phi_U(r,l)$ is the interaction energy between the tip under applied bias, *U,* and the sample polarization, $\Phi_S(r,l)$ is the surface energy of an infinitely thin domain wall, and $\Phi_D(r,l)$ is the depolarization field energy, including the contributions of the surface and bulk bound charges [19, 25, 26]. The domain shape is approximated as a half ellipsoid with the small and large axis equal to *r* and *l*, correspondingly (Fig.1). In this approximation, the domain-size dependent free energy can be represented as a surface in the $(r,l)$ space.

Domain nucleation can be understood from the evolution of free energy Eq. (1) surfaces as a function the bias $U$ applied to the tip. Theoretical analyses performed within the framework of modified point charge approximation of the tip, exact series for sphere-tip interaction energy and capacitance model [25, 26] shows that the stable domain is absent for small biases $U < U_S$, reflecting finiteness of electric field formed below the tip. For larger biases $U_S < U < U_{cr}$, the local minimum $\Phi_{\min} > 0$ appears, corresponding to a metastable domain. The local minimum is separated by a saddle point from the origin. Finally, for $U \geq U_{cr}$, the absolute minimum $\Phi_{\min} < 0$ is achieved corresponding to a thermodynamically stable domain.

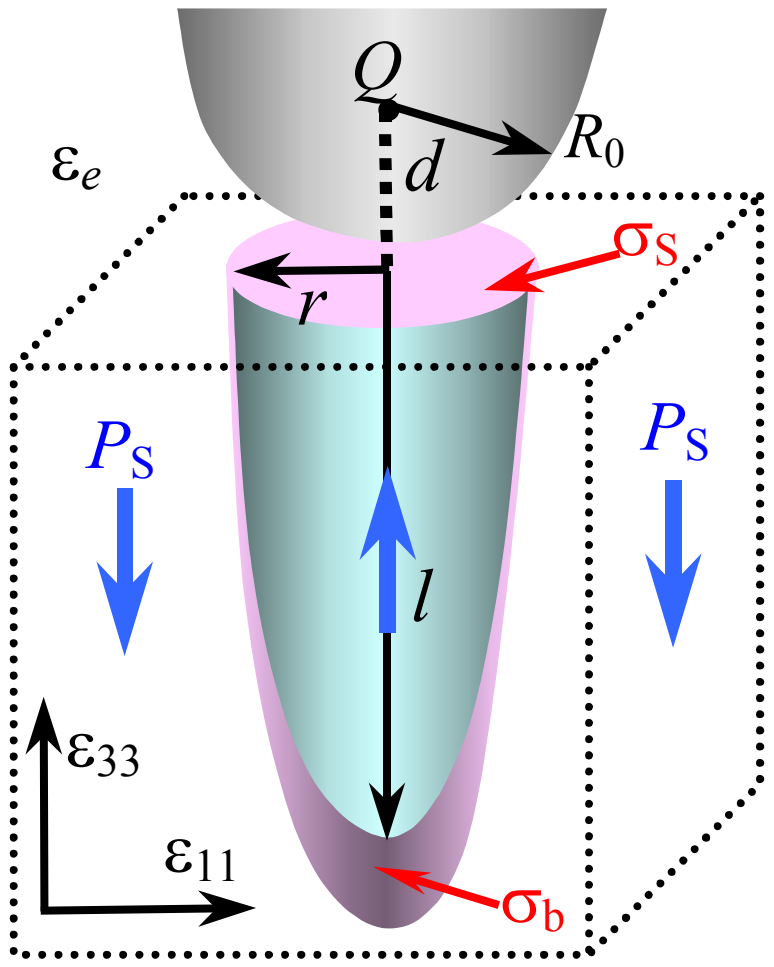

**Fig.1.** Domain formation by the charged PFM tip.

The domain sizes $\{r, l\}$ corresponding to the minima and saddle point, i.e. equilibrium domain size and critical domain size, can be determined as the solution of equations $\partial\Phi(r,l)/\partial r = 0$ and $\partial\Phi(r,l)/\partial l = 0$. Under the additional conditions $\partial^2\Phi(r,l)/\partial r^2 > 0$ and $\partial^2\Phi(r,l)/\partial l^2 > 0$ one obtains the minimum $\{r_{eq}, l_{eq}\}$, i.e. thermodynamically equilibrium domain size. Under the conditions $\partial^2\Phi(r,l)/\partial r^2 < 0$ or $\partial^2\Phi(r,l)/\partial l^2 < 0$ one obtains the saddle point $\{r_S, l_S\}$, i.e. critical domain size. Corresponding energy $\Phi(r_S, l_S) = E_a$ is the nucleation activation energy.

Semi-quantitative analytical results could be obtained from the free energy Pade approximation derived from exact (and extremely cumbersome) formulae derived in Ref. [26] (see also Appendix A for details):

$$\Phi(\tilde{r},\tilde{l}) \approx \frac{-4\tilde{l}\,\tilde{r}^2 f_U U}{\left(\sqrt{\tilde{r}^2+1}+1+2\tilde{l}\right)\left(\sqrt{\tilde{r}^2+1}+1\right)} + 4 f_D \frac{\tilde{r}^3\tilde{l}}{\tilde{l} + 4\kappa\tilde{r}/3\pi(\kappa+\varepsilon_e)} + f_S\,\tilde{l}\,\tilde{r}\,, \qquad (2)$$

where dimensionless domain sizes $\tilde{r} = r/d$ and $\tilde{l} = l/(2d\gamma)$ depend on charge-surface separation $d$ and dielectric anisotropy factor $\gamma = \sqrt{\varepsilon_{33}/\varepsilon_{11}}$. For tip in contact with the surface, the effective point charge separation $d = \varepsilon_e R_0/\kappa$ depends on the dielectric constants of the ambient $\varepsilon_e$ and ferroelectric $\kappa = \sqrt{\varepsilon_{33}\varepsilon_{11}}$; and the tip radius of curvature $R_0$. The interaction constant, $f_U = \pi d^2(P_S - \sigma_S)$, depolarization energy, $f_D = \frac{(P_S - \sigma_S)^2}{3\varepsilon_0(\kappa+\varepsilon_e)} d^3$ and domain wall energy, $f_S = \pi^2 \psi_S d^2 \gamma$, also depend on spontaneous polarization $P_S$, effective screening charge density $\sigma_S$ and domain wall surface energy $\psi_S$.

Note, that in contrast to the early works [18], [19], [25], [26] expansion Eq. (2) is derived for the case when the screening charges captured by the domain apex are spreading over the domain apex allowing for bend bending and field effects in ferroelectric-semiconductor [29] in such a way to reproduce the bound charges distribution $P_S$, but have different effective values at the surface ($\sigma_S$), near the apex of oblate domain (here $\sigma_b \to (P_S + \sigma_S)$), or around the sharp apex of spike-like domain, where $\sigma_b \to 2P_S$ (see Fig.1 and Appendix A for more details).

The evolution of the free energy surface for different voltages without internal screening is depicted in Figs. 2.

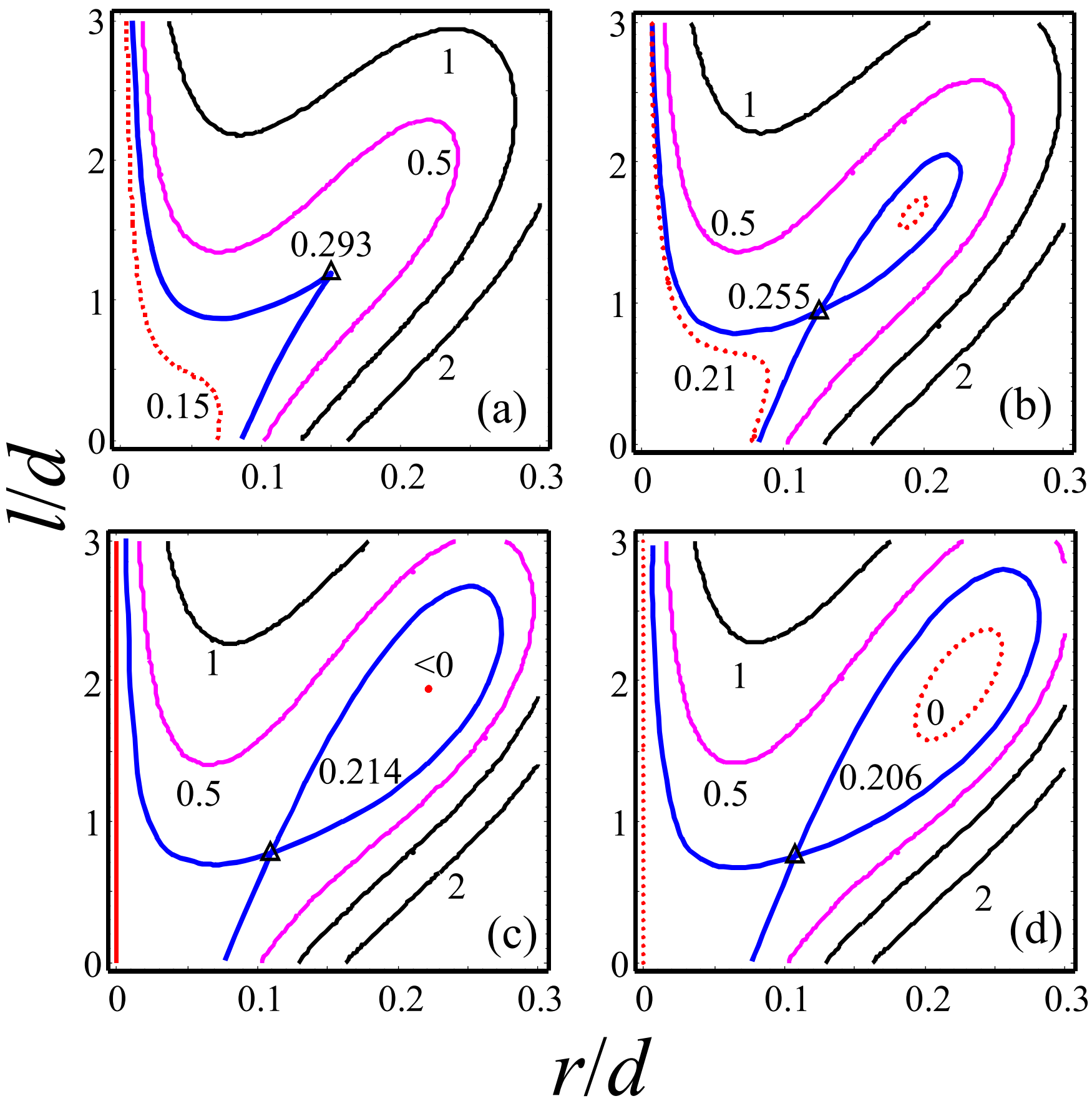

**Fig. 2**. Contour plots of the free energy surface under the voltage increase: (a) instability point - local minimum became to appear ($U = 0.670$ V); (b) saddle point and metastable domain appears ($U = 0.68$ V); (c) transition point - the stable domain appears ($U_{cr} = 0.696$ V); (d) stable domains growth ($U = 0.700$ V). Figures near the contours are free energy values in eV. Triangles denote saddle point (nuclei sizes). Material parameters: $P_S \approx 0.5$ C/m$^2$, domain wall surface energy $\psi_S \approx 25$ mJ/m$^2$, $\kappa = 507$, $\gamma \approx 1$ corresponds to PZT6B ceramics and $d = 4$ nm; $\sigma_S = -P_S$.

The free energy maps at voltages corresponding to onset of stable domain ($U = U_{cr}$) and different screening charge density $\sigma_S$ are shown in Fig.3. Note, that the nucleation bias $U_{cr}$ is almost independent on screening at least at $\sigma_S < +0.5P_S$, whereas the ratio $r(U_{cr})/d$ increases with the screening charge density $\sigma_S$ (compare Figs.3 (a)-(f)).

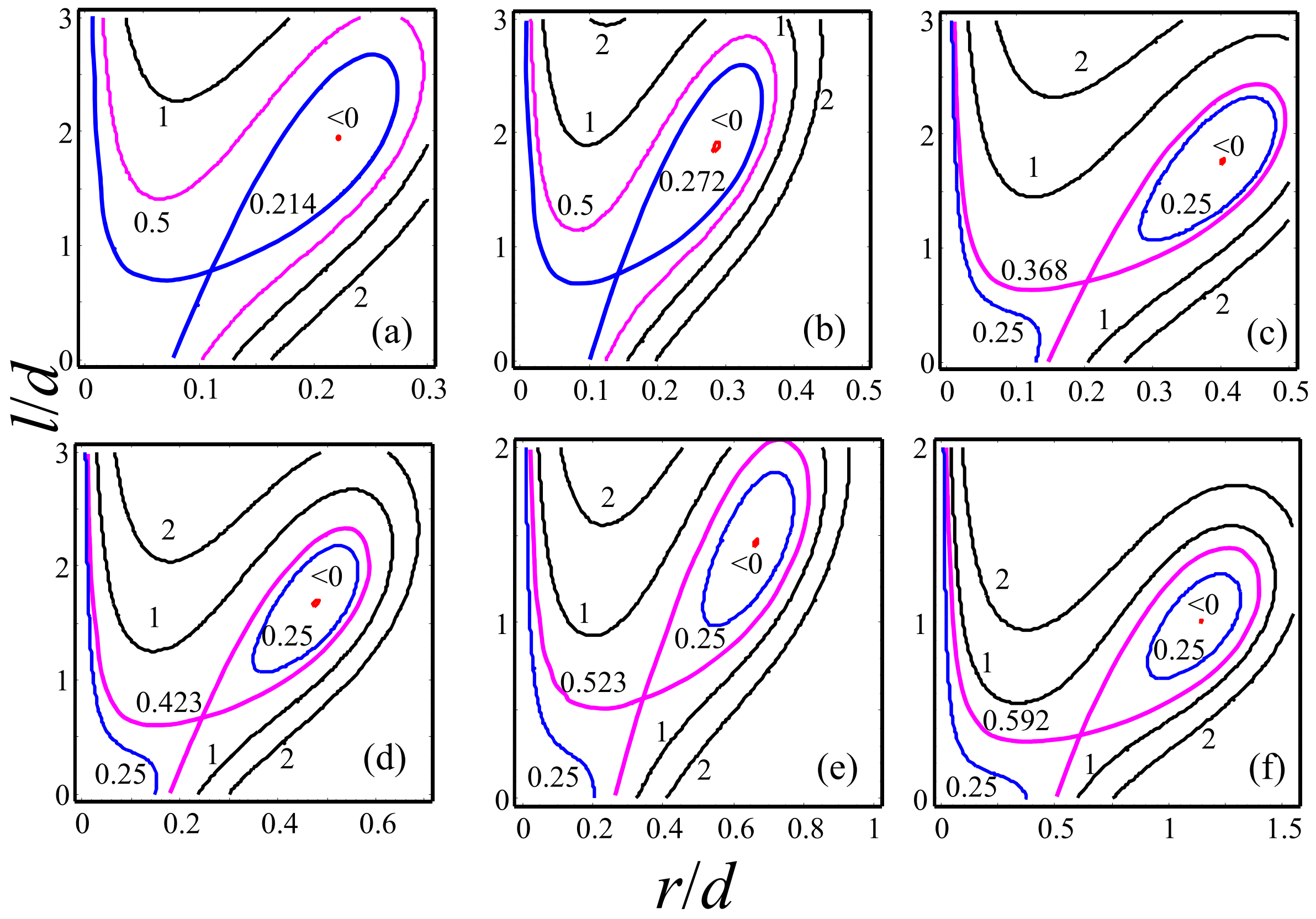

**Fig. 3.** The free energy map at voltages corresponding to onset of thermodynamic switching and different screening charge density $\sigma_S$: (a) $\sigma_S = -P_S$ ($U_{cr} \approx 0.70$ V); (b) $\sigma_S = -0.5P_S$ ($U_{cr} \approx 0.70$ V); (c) $\sigma_S = 0$ ($U_{cr} = 0.72$ V); (d) $\sigma_S = +0.2P_S$ ($U_{cr} = 0.73$ V); (e) $\sigma_S = +0.5P_S$ ($U_{cr} = 0.78$ V); (f) $\sigma_S = +0.8P_S$ ($U_{cr} = 0.98$ V). Figures near the contours are free energy values in eV. Material parameters are given in caption to Fig. 2.

Activation energy $E_a$ vs. applied voltage $U$ for different values of screening charge density $\sigma_S$ is shown in Fig. 4. The nucleation bias $U_{cr}$ is almost independent on $\sigma_S$ at $\sigma_S < +0.5P_S$ in agreement with free energy maps in Figs. 3. At the same time, the bias $U_T$ corresponding to the thermal activation of the domain switching defined as $E_a(U_T) = k_B T/2$ increases with the screening density, $\sigma_S$. In the framework of proposed model the activation barrier for nucleation at $U = U_{cr}$ is minimal for $\sigma_S = -P_S$ (about 0.02eV) and increases up to the 0.9 eV for $\sigma_S = +0.95P_S$ (compare curve 1 - 4).

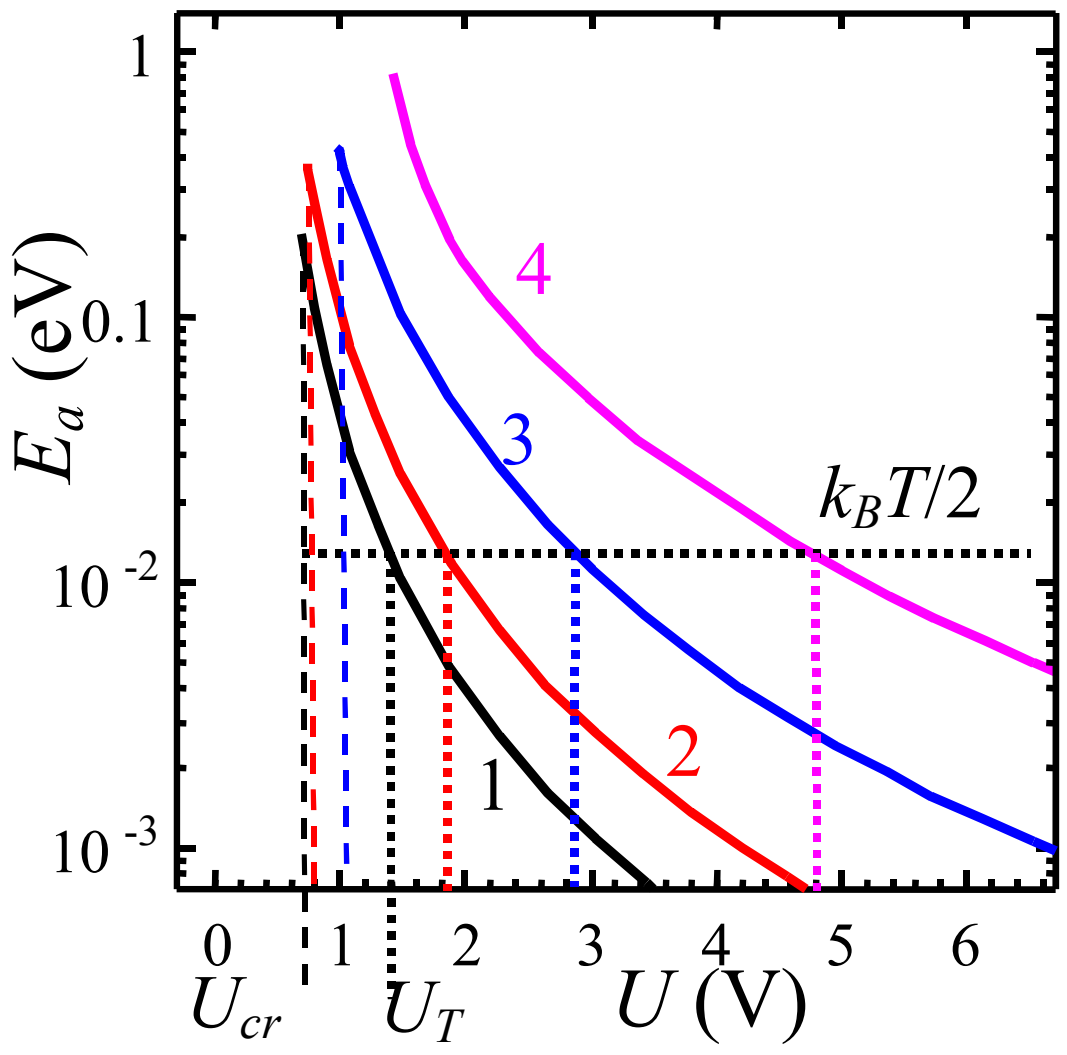

**Fig. 4.** Activation energy $E_a$ vs. applied voltage $U$ for different values of screening charge density $\sigma_S = -P_S;\ +0.2P_S; +0.8P_S; +0.95P_S$ (curves 1, 2, 3, 4). Voltages $U_T = 1.4; 1.9; 2.9; 4.9$ V (vertical dotted lines) correspond to the thermal activation energy $E_a = k_B T/2$ (horizontal dotted line). Material parameters are given in caption to Fig. 2.

Note, that the barrier calculated in the inhomogeneous electric field of the tip is 3-7 orders of magnitude lower than the one calculated by Landauer for the homogeneous electric field.

Hence, thermal activation is possible for materials with good screening at low voltages, while for poor screening surface defects will play the dominant role in switching.

By minimizing the free energy Eq. (2) with respect to dimensionless domain sizes $\tilde{r}$, $\tilde{l}$, we obtain that nontrivial solution ($\tilde{r} \neq 0$, $\tilde{l} \neq 0$) exists only for voltages $U \geq U_S$, where $U_S \approx 3.86\frac{\sqrt{f_D f_S}}{f_U}$ is the bifurcation bias. Namely, for the domain nucleation stage ($r << d$, or equivalently $\tilde{r} < 1$) and its lateral growth at high voltages ($r >> d$) we obtain parametrical dependencies:

$$\tilde{r}(\tilde{l}) = \sqrt{\frac{f_S\left(\tilde{l} + 2\tilde{l}^2\right)}{12 f_D}}, \qquad U(\tilde{l}) = \frac{\sqrt{3 f_D f_S}}{f_U}\cdot\frac{\left(1+\tilde{l}\right)^2}{\sqrt{\tilde{l} + 2\tilde{l}^2}}, \qquad \tilde{r} << 1 \tag{3a}$$

$$\tilde{r}(\tilde{l}) = \sqrt{\frac{f_S}{12 f_D}}\cdot\tilde{l}, \qquad \tilde{l}(U) = 2\sqrt{\frac{f_U U}{f_S}} - 1, \qquad \tilde{r} >> 1 \tag{3b}$$

Parametrical dependence for $U(\tilde{l})$ given by Eq. (3a) allows asymptotic expansions to be derived as

$$U(\tilde{l}) = \frac{\sqrt{3 f_D f_S}}{f_U}\cdot\frac{\left(1+\tilde{l}\right)^2}{\sqrt{\tilde{l} + 2\tilde{l}^2}} \approx \frac{\sqrt{3 f_D f_S}}{f_U}\begin{cases}\frac{1}{\sqrt{\tilde{l}}} + \sqrt{\tilde{l}} + O\left(\sqrt{\tilde{l}}^3\right), & \tilde{l} << 1 \\ \frac{\tilde{l}}{\sqrt{2}} + \frac{7}{4\sqrt{2}} + \frac{19}{32\sqrt{2}\cdot\tilde{l}}, & \tilde{l} >> 1\end{cases} \tag{4}$$

Parametric dependences $l(U)$ and $r(U)$ given by Eq.(3a) are depicted in Fig. 5. From the graph in Fig. 5 it is clear that Eq. (4) for small $\tilde{l}$ gives the asymptotic representation for the saddle point coordinates $\{\tilde{r}_S, \tilde{l}_S\}$ and corresponding activation energy $E_a = \Phi(r_S, l_S)$. Notably, $\tilde{r}_S << 1$ justifying the validity of Eq.(3a) and self-consistence of calculations.

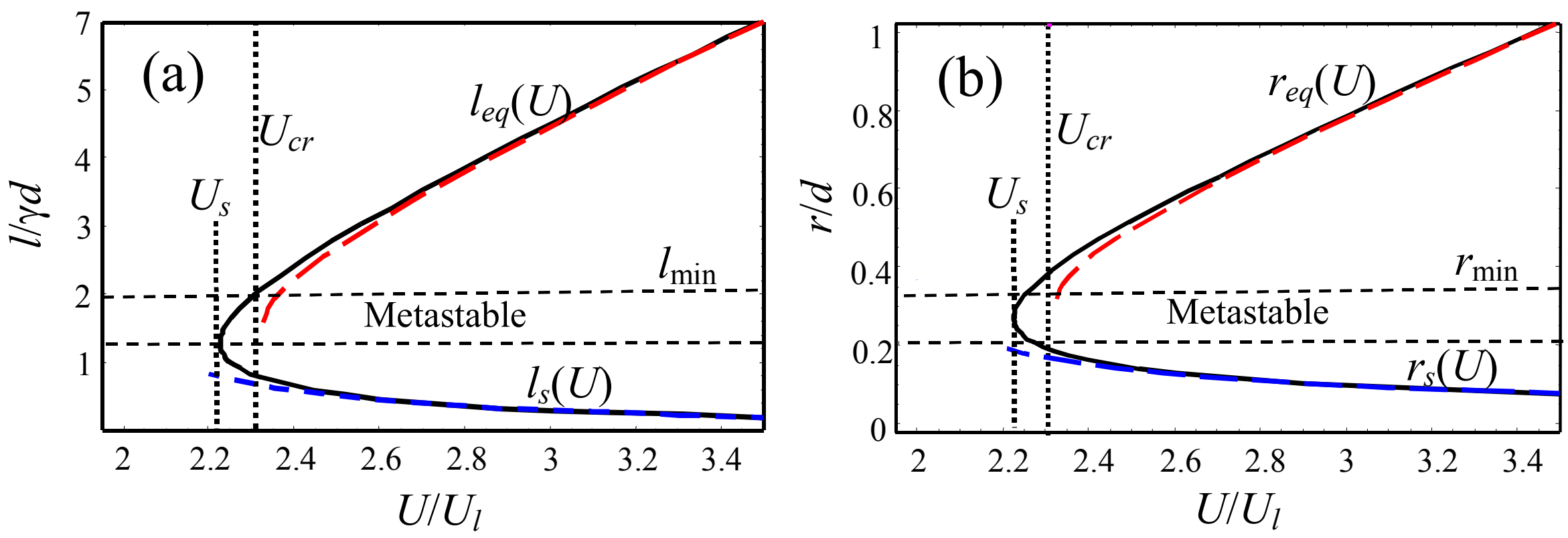

**Fig.5.** Parametric dependence Eq. (4) of domain length $l/d\gamma$ (a) and radius $r/d$ (b) via applied bias $U/U_l$, where $U_l = \sqrt{3 f_D f_S}/f_U$, $\sqrt{f_S/12 f_D} = 0.05$. Dashed curves represent asymptotic expansions.

In the limit $r < d$, corresponding to the small size of the domain compared to the tip radius, we used approximate interaction energy $\frac{-4\tilde{l}\,\tilde{r}^2 f_U U}{\left(\sqrt{\tilde{r}^2+1}+1+2\tilde{l}\right)\left(\sqrt{\tilde{r}^2+1}+1\right)} \approx -f_U U \frac{2\tilde{l}\,\tilde{r}^2}{\left(1+\tilde{l}\right)}$ and depolarization energy $4 f_D \frac{\tilde{r}^3 \tilde{l}}{\tilde{l} + 4\kappa\tilde{r}/3\pi(\kappa+\varepsilon_e)} \approx 4 f_D \tilde{r}^3$ to obtain analytical expressions for the nucleus (saddle) and critical points parameters. Corresponding bias dependences of nucleus domain parameters are:

$$l_S(U) = \frac{12 f_D f_S \gamma d}{f_U U\left(f_U U + \sqrt{(f_U U)^2 - 12 f_D f_S}\right) - 6 f_D f_S} \approx \frac{6 f_D f_S}{(f_U U)^2}\gamma d, \tag{5a}$$

$$r_S(U) = \frac{f_S f_U U d}{f_U U\left(f_U U + \sqrt{(f_U U)^2 - 12 f_D f_S}\right) - 6 f_D f_S} \approx \frac{f_S}{2 f_U U} d, \tag{5b}$$

$$E_a(U) = \frac{4 f_S^3 f_D f_U U\left((f_U U)^2 - 9 f_D\, f_S\right)}{\left(f_U U\left(f_U U + \sqrt{(f_U U)^2 - 12 f_D\, f_S}\right) - 6 f_D\, f_S\right)^3} \approx \frac{f_D}{2}\left(\frac{f_S}{f_U U}\right)^3. \quad \text{(5c)}$$

Using Eqs. (3-4) and condition $\Phi\left(\tilde{r}_{\min}, \tilde{l}_{\min}\right) = 0$ we derived the following expressions for critical voltage $U_{cr}$ and sizes $r_{\min}$, $l_{\min}$

$$U_{cr} = 4\frac{\sqrt{f_D f_S}}{f_U}, \qquad l_{\min} = 2\gamma\, d, \qquad \frac{r_{\min}}{d} = \frac{1}{2}\sqrt{\frac{f_S}{f_D}}. \quad \text{(6)}$$

Note, that the bifurcation bias $U_S$ is numerically close to the critical bias $U_{cr}$, i.e. $U_S / U_{cr} = 0.965$. Therefore, the bias corresponding to the metastability of domain and thermodynamic stability are close and are unlikely to be differentiated experimentally.

It should be noted that critical voltage $U_{cr}$ given by Eq. (6) is independent on the screening charge density $\sigma_S$, whereas corresponding domain size is dependent on screening mechanism, $r_{\min}/d \sim (P_S - \sigma_S)^{-1}$. The same is true for activation energy, $E_a / f_S \sim (U_{cr}/4U)^3\, r_{\min}/d$ (compare with Figs.3-4). The combination of material constants $\sqrt{f_S / f_D}$ that determines the domain critical radius and its aspect ratio $r_{\min}/d$ is closely related with the domain wall thickness $w \cong \varepsilon_0 \kappa \psi_S / P_S^2$, predicted by the phenomenological Ginsburg-Landau-Devonshire theory (see e.g. Ref. [30]), namely $\sqrt{f_S/f_D} \sim \sqrt{w/d}$. Since $w$ usually is about of few lattice constants and $d$ is from several to tens of nanometers, $\sqrt{f_S/f_D}$ is smaller than unity at $\sigma_S \sim -P_S$. However at $\sigma_S \to P_S$ parameter $f_D$ tends to zero and $\sqrt{f_S/f_D}$ can be much larger than $\sqrt{w/d}$. Since the form of the domain is determined by the ratio $r/l = \sqrt{\frac{f_S}{3 f_D}} \cdot \frac{1}{4\gamma}$, for $\sigma_S \to P_S$ domain changes its form from prolate to oblate.

Using Eqs. (3, 6) we obtained approximate expressions for the bias dependences of the equilibrium domain size $\{r_{eq}, l_{eq}\}$ as:

$$r_{eq}(U) \approx r_{\min}\left(1+2\sqrt{\frac{2d}{3r_{\min}}\left(\frac{U}{U_{cr}}-1\right)}\right), \quad l_{eq}(U) \approx l_{\min}\left(1+2\sqrt{\frac{2d}{r_{\min}}\left(\frac{U}{U_{cr}}-1\right)}\right). \qquad (7)$$

Hence, for $\sigma_S \sim -P_S\, r_{eq}/l_{eq} \sim \sqrt{w/d}$ at $U >> U_{cr}$.

### 3. Effective piezoelectric response

The self-consistent 3D model for piezoelectric hysteresis loop formation is based on (a) deriving the main parameters of domain nucleation in semi-infinite material and (b) establishing the relationships between domain parameters, tip characteristics ($r$, $l$ and $d$) and piezoresponse signal using decoupled Green's function theory. For data interpretation in realistic experiment, the third step (c) is calibration of the probe geometry using appropriate standard (e.g. domain wall width).

Simplified expression for the effective piezoelectric response $d_{33}^{eff}$ measured in the center of a cylindrical domain of radius $r$ has been previously derived as [31]:

$$d_{33}^{eff}(r) \approx -\left(\frac{3d_{33}^{*}}{4}\left(1-\frac{2}{1+\pi d/8r}\right)+\frac{d_{15}}{4}\left(1-\frac{2}{1+3\pi d/8r}\right)\right). \qquad (8)$$

Here $r$ is the domain radius and $d$ is charge-surface distance; $d_{33}^{*} = d_{33} + (1/3 + 4\nu/3)d_{31}$, $d_{ij}$ is the piezoelectric tensor components, $\nu$ is the Poisson ratio and materials is assumed to be close to dielectric isotropy, $\gamma \approx 1$. The combination of Eqs. (7, 8) allows self-consistent analysis of piezoresponse bias dependence. Typical piezoelectric response $d_{33}^{eff}$ hysteresis loops calculated in weak pinning limit [25] for different screening conditions are depicted in Figs.6; appropriate activation energy $E_a$ bias dependences are shown in Fig.4.

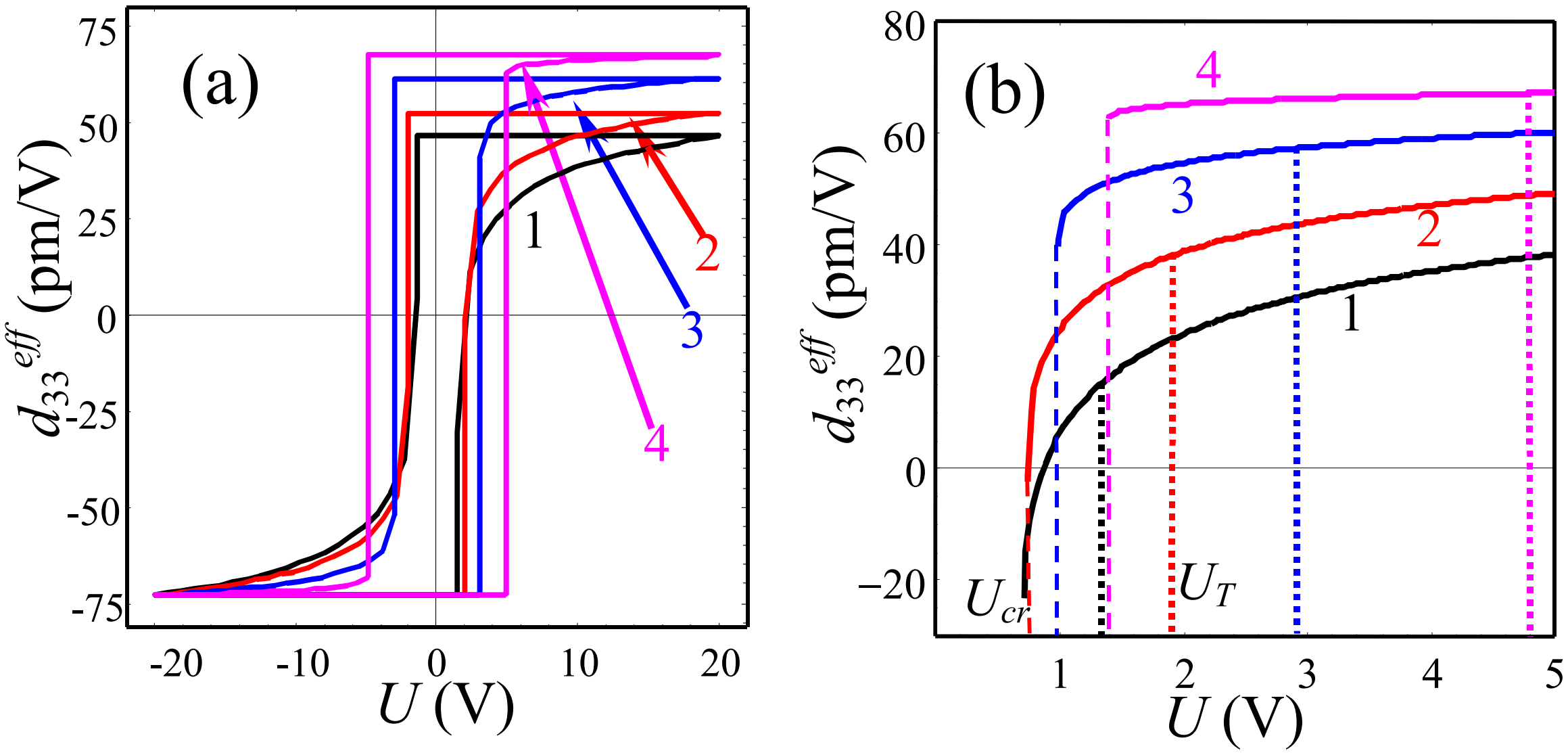

**Fig. 6.** Piezoelectric response $d_{33}^{eff}$ hysteresis loops for thermal domain nucleation calculated in weak pinning limit (a) and thermodynamic forward branch (b) calculated for $r_{\min}/d \approx 0.5(1-\sigma_S/P_S)^{-1}$ under the different values of screening charge density $\sigma_S = -P_S$; $+0.2P_S$;$+0.8P_S$;$+0.95P_S$ (curves 1, 2, 3, 4). Dashed lines in (b) are critical voltage $U_T$, whereas the dotted ones correspond to the thermal activation at voltages $U_T$. Piezo-modules $d_{ij}$ correspond to PZT6B ceramics.

Depending on the switching rate, the initial domain nucleation occurs at voltages $U_{cr} \le U \le U_T$ (forward branch). Then domain sizes increase under the further voltage increase up to $U = +U_{\max}$. On the reverse branch of the hysteresis loop, the domain does not shrink. Rather, the domain wall is pinned by the lattice and defects, maintaining constant size.[32] The inverted domain appeared only at $-U_T \le U \le -U_{cr}$. A sufficiently large domain acts as new matrix for the inverted one, appearing just below the tip. The inverted domain size increases with further voltage decrease. At the point $U = -U_{\max}$, the domain walls annihilate and the system returns to the initial state. It is reasonable to suppose that fast switching rate corresponds to the nucleation at $U = \pm U_T$, whereas the slow ones to $U = \pm U_{cr}$.

Note that in this example the vertical asymmetry of the loop (downward shift) follows from the fact that the response of the nested domains forming when the switching is incomplete differs from the single one. Domain walls annihilate at maximal negative voltage $U = -U_{\max}$. Then response coincides with the one from the initial state. The loop vertical asymmetry decreases under the maximal voltage increase [26] and/or $\sigma_S$ increase, namely the loop that corresponds to $\sigma_S = -P_S$ is strongly asymmetrical, whereas the loop that corresponds to $\sigma_S = +0.95P_S$ becomes almost symmetrical.

At the same time, screening clearly affects the saturation rate. For strong screening, the nucleating domain is small and increases gradually with the bias, resulting in slow gradual saturation of the hysteresis loop. For poor screening, the nucleating domain is large, and the loop saturates much faster. However, corresponding activation energies are much higher, indicative of the increased role of defects on nucleation process. Hence, for poor screening the hysteresis loops are sharper and are expected to be less reproducible from point to point. These observations correspond to experimental observation of slowly saturating hysteresis loops for materials such as PZT, and fast saturation for $BiFeO_3$.

## IV. Conclusions

To summarize, in this manuscript we derived the closed-form analytical expressions for the nucleation threshold and bias dependence of critical nucleus size and activation energy and equilibrium domain size in PFM experiment. The role of surface and bulk screening on these processes is established and in particular it is shown that nucleation bias is independent on screening process. For good screening, thermal activation is possible due to low activation energies, while for poor screening the defects can strongly affect the nucleation process, similar to macroscopic switching.

We have shown that screening charges control both activation energy value and hysteresis loop saturation rate, the latter strongly increases at $\sigma_S \rightarrow +P_S$ . Furthermore, sharper hysteresis loops are anticipated for unscreened surfaces. This analysis establishes the guidelines for the interpretation of PFS data on ferroelectric surfaces in different environments affecting screening mechanisms.

Research sponsored by the Division of Materials Sciences and Engineering, Office of Basic Energy Sciences, U.S. Department of Energy, under Contract DE-AC05-00OR22725 with Oak Ridge National Laboratory, managed and operated by UT-Battelle, LLC (SVK).

## Appendix A

Depending on the domain length screening charge $\sigma_b$ concentrated near the domain wall has different values at $r << l$ (oblate shape) and $r >> l$ (prolate shape) allowing for the charge conservation for oblate domain (initial surface screening charge is $\pm P_S$ depending on the polarization orientation split into the $-\sigma_S$ at surface and $\sigma_b \to (P_S + \sigma_S)$ in the bulk) and the concentration of depolarizing electric field near the apex of prolate domain (field sharply increases proportionally to $l^2/r^2$ and so $\sigma_b \to 2P_S$). Thus we assume that:

$$\sigma_b = \begin{cases} P_S + \sigma_S, & r >> l, \\ 2P_S, & r << l. \end{cases} \quad \text{(A.1)}$$

Note, that the condition $\sigma_b \to 2P_S$ corresponds to the domain breakdown reported in Ref.[14]. Using Pade approximations for the interaction energy:

$$\Phi_U(r,l) \approx 2\pi d^2 U \left( \frac{(\sigma_S - P_S) r^2}{\sqrt{r^2 + d^2} + d} + \frac{(2P_S - \sigma_b) r^2}{\sqrt{r^2 + d^2} + d + l/\gamma} \right) = \\ = 2\pi d^2 U \begin{cases} \dfrac{(\sigma_S - P_S) r^2}{\sqrt{r^2 + d^2} + d}, & r << l \\ \dfrac{(\sigma_S - P_S) r^2}{\sqrt{r^2 + d^2} + d} + \dfrac{(P_S - \sigma_S) r^2}{\sqrt{r^2 + d^2} + d + l/\gamma}, & r >> l \end{cases} \quad \text{(A.2)}$$

Joining of Eq.(A.2) leads to the following Pade approximation:

$$\Phi_U(r,l) \approx \frac{2\pi d^2 U (\sigma_S - P_S) r^2 \, l/\gamma}{\left(\sqrt{r^2 + d^2} + d\right)\left(\sqrt{r^2 + d^2} + d + l/\gamma\right)} \quad \text{(A.3)}$$

Using Eq.(A.1) we derive the following expression for the upper estimation of the total depolarization field energy

$$\Phi_D(r,l) \le \begin{pmatrix} \frac{4(P_S-\sigma_S)^2}{3\varepsilon_0(\kappa+\varepsilon_e)} r^3 + \frac{4(2P_S-\sigma_b)^2}{6\varepsilon_0\kappa} r^3 \left(1+\frac{\kappa-\varepsilon_e}{\kappa+\varepsilon_e}\mathrm{Int}\left(\frac{2l}{\gamma r}\right)\right) - \\ -\frac{8(P_S-\sigma_S)(2P_S-\sigma_b)}{3\varepsilon_0(\kappa+\varepsilon_e)} r^3\, \mathrm{Int}\left(\frac{l}{\gamma r}\right) \end{pmatrix} \quad \text{(A.4)}$$

Where the first term is related the surface charges (bond and screening), the second one is caused by the bulk bond charges and their images in the sample surface, and the third is determined by the interaction between the surface and bulk charges. One parametric function $\mathrm{Int}(x)$ is introduced as:

$$\mathrm{Int}(\lambda) = \int_0^\infty \frac{3\pi}{4}\left(\frac{\mathrm{J}_1(x)}{x}\right)^2 \exp(-\lambda x)dx = \begin{cases} 1-\frac{3\pi}{8}\lambda, & \lambda << 1; \\ \frac{3\pi}{16}\frac{1}{\lambda}, & \lambda >> 1; \end{cases} \quad \text{(A.5)}$$

Joining of Eqs.(A.4-5) and model (A.1) leads to the following Pade approximation for depolarization energy:

$$\Phi_D(r,l) = \begin{Bmatrix} \frac{\pi(P_S-\sigma_S)^2}{2\varepsilon_0\kappa\gamma} r^2 l, & l << \gamma r, \\ \frac{4(P_S-\sigma_S)^2}{3\varepsilon_0(\kappa+\varepsilon_e)} r^3, & l >> \gamma r. \end{Bmatrix} \approx \frac{4(P_S-\sigma_S)^2}{3\varepsilon_0(\kappa+\varepsilon_e)} r^3 \frac{l}{l+\frac{8\gamma r\kappa}{3\pi(\kappa+\varepsilon_e)}} \quad \text{(A.6)}$$

Note, that in the case of a plane upper electrode (Landauer problem) instead of a localized conductive tip, Pade approximation (A.6) gives the upper estimation of Landauer depolarization energy, i.e. $\Phi_D(r,l) \ge \Phi_{DL}(r,l)$, where $\Phi_{DL}(r,l) \cong \frac{2(2P_S-\sigma_b)^2}{3\varepsilon_0\kappa}\frac{r^4}{l}\ln\left(\frac{2l}{r\gamma}\right)$ ($r < l$).